\documentclass[prl,twocolumn,showpacs,amsmath,amssymb]{revtex4}
\usepackage[dvips]{graphicx}
\usepackage{latexsym}
\usepackage{amssymb}
\usepackage{amsmath}
\usepackage{amsbsy}
\usepackage{psfrag}

\def\b{\beta}

\def\d{\delta}
\def\GG{\Gamma}

\def \e{\varepsilon}

\def\h{\hbar}
\def \oo {\omega}
\def\OO {\Omega}

\def\Tr{\mbox{tr}\,}
\def\F{{\mathcal F}}

\def\S{{\mathcal S}}

\def\m{\mu}
\def\DD{\partial}
\def\dwell{\tau_{\rm d}}

\def\G {\Gamma}

\def\Rft {R_{{\rm 4t}}}
\def\Rtt {R_{{\rm 2t}}}

\def\la {\langle}
\def\ra {\rangle}

\def \etal{{\it et al.}}

\def\RC {\tau_{\rm RC}}
\def\Var{\mbox {Var }}

\def\eg{{\it e.g. }}

 \psfrag{f}{$f$} \psfrag{r}{$r$}
\psfrag{V0}{$V_0$} \psfrag{Vtilde}{$\tilde V$} \psfrag{eta}{$\eta$}
\psfrag{N1}{$N_1$}\psfrag{N2}{$N_2$}\psfrag{C0}{$C_{g0}$}\psfrag{C1}{$C_1$}
\psfrag{C2}{$C_2$}
\psfrag{Fa}{$\Phi_a$}\psfrag{Fd}{$\Phi_d$}\psfrag{F}{$\Phi$}\psfrag{otimes}{$\otimes$}\psfrag{S}{${\cal
S}_{N\times N}$}\psfrag{U}{${\cal U}_{M\times M}$}\psfrag{NxN}{$
$}\psfrag{MxM}{$ $}\psfrag{G}{$\cal
G$}\psfrag{ho}{$\hbar\oo$}\psfrag{Cparas}{$C_{\rm
stray}$}\psfrag{Vp}{$V_{\rm p}$}\psfrag{Cpump}{$C_{\rm p}$}
\begin{document}

\title{Dynamics of four- versus two-terminal transport through chaotic quantum cavities}
\date{\today}
\author{M.~L.~Polianski $^{1,2}$}\author{M. B\"uttiker$^3$}
\affiliation{$^1$ Niels Bohr Institute,
          NBIA,
          Blegdamsvej 17,
          DK-2100 Copenhagen
          Denmark
          \\ $^2$ Laboratoire de Physique des Solides,
          Universit\'e Paris-Sud, 91405 Orsay, France
          \\ $^3$ D\'epartement de
Physique Th\'eorique, Universit\'e de Gen\`eve, CH-1211 Gen\`eve 4,
Switzerland}
\begin{abstract}
We consider multi-terminal mesoscopic transport through a
well-conducting chaotic quantum cavity using random matrix theory.
Four-probe resistance vanishes on the average and is not affected by
weak localization. Its fluctuations are given by a single expression
valid for arbitrary temperature, ac frequency, non-ideal coupling of
the contacts, and in the presence of floating probes; surprisingly,
they are governed by the dwell time only. In contrast, the two-probe
transport additionally depends on the RC-time, which is interpreted
as a property of the measurement scheme. We also predict a universal
mesoscopic distribution of the
 phase of transmitted voltage in an ac experiment.
\end{abstract}
\pacs{73.23.-b,05.60.Gg,73.63.Kv}


 \maketitle

\begin{figure}[b]
\centering\includegraphics[width=8.5cm]{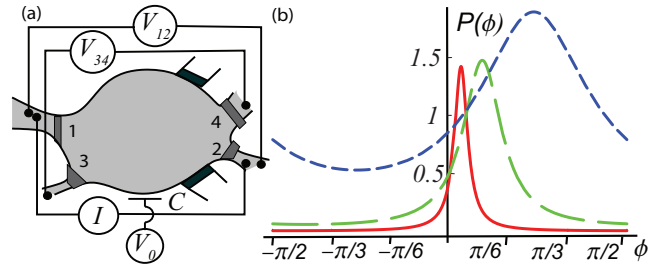}
\caption{(a) A multi-terminal chaotic cavity with a gate
(capacitance $C$). A current through source-drain contacts 1,2 and a
voltage drop between 3,4 (or 1,2) determine the four-probe
$\Rft=V_{34}/I$ (or two-probe $\Rtt=V_{12}/I$) resistances.
Additional floating probes draw no current. (b) The mesoscopic
distribution $P(\phi)$ (rescaled by $\times\oo\dwell$) of phase
$\phi$ of the transmitted ac voltage as a function of frequency, for
$\oo\dwell=0.1,0.3,1$, full, long- and short-dashed, respectively. }
\label{FIG:setup}
\end{figure}

{\it Introduction.} Mesoscopic transport and its sample-to-sample
fluctuations are fundamentally important for understanding quantum
effects in electronic transfer through nano-structures. However,
when a voltage is measured between the source and drain in a
two-terminal geometry, a big classical contribution to transport
shadows quantum effects of order $e^2/h$ \cite{imry2002imp,MPS}. The
visibility of interference effects can be enhanced by diminishing
the width of the contacts to reservoirs, but this also leads to
Coulomb blockade of electrons\,\cite{MPS}. Another approach is to
keep the contacts well-conducting and apply a multi-terminal scheme
with a voltage through the sample measured by two {\it additional}
contacts, see Fig.\,\ref{FIG:setup}(a). The thus defined
four-terminal resistance $\Rft$, well-known to experimentalists, has
some benefits over the two-terminal one, $\Rtt$, which is more
popular among theoreticians.

The central advantage of $\Rft$ is to exclude the contact
properties. Indeed, classical resistances are local, and a
four-terminal scheme removes the effect of the contacts and
therefore probes the sample's properties only
\cite{vanderpauw:1958}. Quantum mechanically resistance is not
local, and this scheme does reduce the role of contacts but they
still affect the measurement. Nevertheless even for a coherent
sample there are of course profound differences between $\Rtt$ and
$\Rft$. The former determines Joule heating and is therefore always
positive and symmetric to the magnetic field inversion. In contrast,
the latter need not be
positive\,\cite{Buttiker4terminal:1986,ButtikerIBM:1988} and
experiments demonstrate its fluctuations around zero
\cite{DePicciotto:2001,Gao:2005,Gunnarsson:2008} and magnetic field
asymmetry \cite{Benoit:1986}.

Non-universal fluctuations in $\Rft$ were widely investigated
in metallic diffusive samples, where small quantum effects on top of
a large classical average depend on the probe
locations\,\cite{MaLee:1987,SpivakZyuzin:MPS,BarangerStoneDiVincenzo:1988,
KaneLeeDiV:1988,Iida:1991,Schmidt:1993}. On the other hand, in a
chaotic cavity\,\cite{Beenakker:1997,Aleiner:2002} the exact
positions of probes are irrelevant, the classic voltage drop
vanishes, and an experiment would measure quantum effects directly.
Remarkably, $\Rft$ in this generic geometry remained unexplored
except for a recent initial experiment \cite{Lerescu:2007}.

What could one expect from $\Rft$ in a quantum dot? Its dc
two-terminal conductance shows a weak-localization correction (WL)
$\sim e^2/h$ due to an enhanced return probability for an electron,
and universal conductance fluctuations (UCF) $\sim (e^2/h)^2$; a
magnetic field destroys WL and reduces UCF by a factor
2\,\cite{Beenakker:1997}. These results for ballistic quantum point
contacts (QPCs) can be generalized to include {\it either} (i) a
dephasing (floating) probe coupling\,\cite{Brouwer:1997}, or (ii)
non-ideal QPCs\,\cite{BrouwerBeenakker:1996}, or (iii) a
low-frequency ac setup\,\cite{BrouwerButtiker:1997}. In (iii) the
averaged conductance\,\cite{BrouwerButtiker:1997} and
shot-noise\,\cite{polianski:2005} in their leading order depend only
on the charge relaxation time (RC-time) $\RC$, sensitive to the
Coulomb interaction energy $e^2/2C$ of the capacitor.
 Its appearance is natural: at high
frequencies $\oo\gtrsim 1/\RC$ the capacitor conducts better than
the contacts. However, WL and UCF, as well as the third current
cumulant, additionally depend on the dwell time $\dwell$ an electron
typically spends in the
dot\,\cite{BrouwerButtiker:1997,Nagaev:2004,Petitjean:2009}. As a
result, even in linear transport there is no unique time-scale for
dispersion of various quantities.

At first one expects $\Rft$ to share the features of two-terminal WL
and UCF. However, these expectations are wrong: $\Rft$ does not have
weak localization, and its fluctuations are insensitive to magnetic
field. The differences are even greater when the dynamics of
resistance fluctuations is compared: unlike $\Rtt$ governed both by
$\dwell$ and $\RC$, $\Rft$ turns out to depend only on $\dwell$.
Surprisingly, fluctuations of $\Rft$ are much simpler than UCF, and
a single formula covers the regimes (i-iii) for arbitrary floating
probes, imperfections in contacts, and frequency $\oo$.

{\it Universal resistance fluctuations.} This Letter compares
statistics of two- and four-terminal resistances in quantum dots
with non-ideal coupling to reservoirs and probes, biased by ac
voltages, see Fig.\,\ref{FIG:setup}(a). To be specific let the
current $I$ flow only through contacts 1,2 and measure the voltage
drop between 3,4 (or 1,2) to find $\Rft=V_{34}/I$ (or
$\Rtt=V_{12}/I$). Each contact $j$ with $N_j$ orbital channels is
specified by a diagonal matrix $\GG_j$ of its channel transmissions,
and we assume good dimensionless conductances of the probes 1--4,
$\Tr\GG_j\gg 1$. At arbitrary temperature $T$ and measurement
frequency $\oo$, in first two orders in the total QPC conductance
$\Tr\GG$, the ensemble averaged resistances are given by
\begin{eqnarray}\label{eq:acRft}
\la\Rft\ra &=& 0,\\
\label{eq:acRtt}
 \la\Rtt\ra
&=&\frac{h}{2e^2}\frac{\Tr\GG_1+\Tr\GG_2}{\Tr\GG_1\Tr\GG_2}
\left[1+\frac{\d_{\b 1}}{\Tr\GG(1-i\oo\dwell)}\right].
\end{eqnarray}
Vanishing $\la\Rft\ra$ is not sensitive to the presence of
time-reversal symmetry (TRS), $\b=1$, or its breaking by magnetic
field, $\b=2$. In contrast, $\la\Rtt\ra$ is corrected by WL
depending only on $\dwell$. Similarly, TRS does (not) affect
fluctuations of $\Rtt(\Rft)$ around their averages:
\begin{eqnarray}
\Var\Rft&=&
\left(\frac{h}{2e^2\Tr\GG}\right)^2\textstyle\sum_{\substack{i=1,2\\v=3,4}}\displaystyle\frac{\Tr\Gamma_i^2}{\Tr^2\Gamma_i}
\frac{\Tr\Gamma_v^2}{\Tr^2\Gamma_v}\int_{\oo\oo},\label{eq:Rmainac}\\
\Var\Rtt&=&\frac{2}{ \b}\left(\frac{h}{2e^2}\frac{N_1+N_2}{N_1
N_2 N}\right)^2\frac{1-2i\oo\RC} {(1-i\oo\RC)^2}\int_{\oo\oo}\label{eq:acUCF},\\
\int_{\oo\OO} &\equiv&\int_0^{+\infty} \left(\frac{2\pi T\sin
\oo\tau/2}{\h\oo\sinh \pi T\tau/\h}\right)^2 \frac{e^{\tau(i\OO-
1/\dwell)}d\tau} {(1-i\OO\dwell)\dwell}.
\label{eq:int}
\end{eqnarray}
Fluctuations of $\Rft$, Eq.\,(\ref{eq:Rmainac}), the main result of
this paper, are valid for any couplings of floating probes, and
imperfections in the contacts, while Eq.\,(\ref{eq:acUCF}) is
derived only for ballistic QPCs in a multi-terminal dot.

In an effectively zero-dimensional dot universal Eqs. (1--4) do not
depend on disorder and contact positions. It is interesting that for
conductance there are no results of comparable generality for WL and
UCF . We conclude that both weak-localization corrections to the
two-terminal resistance, Eq.\,(\ref{eq:acRtt}), and fluctuations of
the four-terminal one, Eq.\,(\ref{eq:Rmainac}), are {\it completely
insensitive at all frequencies} to the RC-time and thus to Coulomb
interactions. We next explain how these results are obtained,  and
discuss the dc fluctuations of resistances. Later we generalize our
results onto an ac setup and discuss the role of the chosen
measurement scheme in fluctuations. We calculate the universal
distribution of the phase of transmitted voltage in a four-terminal
ac experiment.

{\it The calculation.} We consider a chaotic quantum dot in a
multi-probe setup with $M\geq 4$ contacts and a gate with
capacitance $C$, see Fig.\,\ref{FIG:setup} (a). Each contact
$j=1,...,M$ leading to a reservoir with a voltage $V_{j\oo}$ at the
frequency $\oo$ is characterized by a set of transmission values
$\GG_j$ of its (imperfect) channels. This diffusive/ballistic dot is
in the universal regime, when direct trajectories are absent, and
chaos validates random matrix theory
(RMT)\,\cite{Beenakker:1997,Aleiner:2002}. The Thouless energy of a
closed dot is much larger than the temperature $T$, mean level
spacing $\Delta=2\pi\h^2/(m\mbox{ Area})$, the excitation energy
$\hbar\oo$ in the ac experiment, and escape rate $\Tr\GG
\cdot\Delta/2\pi$. The total coupling to environment is good,
$\Tr\GG\gg 1$, and it specifies the electronic dwell-time
$\dwell=h/(\Tr\GG\Delta)$ in the dot. The leading Coulomb
interaction effect is accounted for by a uniform self-consistent
potential of the dot, found from charge
conservation\,\cite{ButtikerPretre:1993PRL,BrouwerButtiker:1997}.
The terms leading to Coulomb blockade are unimportant due to
$1/\Tr\GG\ll 1$. We use RMT and diagrammatic technique with a small
parameter $1/\Tr\GG\ll 1$ for an energy-dependent scattering matrix
$\S(\e)$ of a dot with imperfect
contacts\,\cite{Brouwer:1995,BrouwerBeenakker:1996,
PolianskiJPHYSA:2003} to find transport statistics. Further we take
$h=e=1$ and consider spin-degenerate electrons, $\nu_s=2$, the role
of spin-orbit in $\Rft$ being discussed
elsewhere\,\cite{PolianskiIFE:2009}.

Linear transport coefficients are found from the screened
frequency-dependent degenerate $(M+1)\times(M+1)$ conductance matrix
$g_{ij,\oo}^s=\DD I_{i\oo}/\DD V_{j\oo}$, where
\begin{eqnarray}
g_{ij,\oo}^s &=& g_{ij,\oo}-\textstyle\sum_{kl}
g_{il,\oo}g_{kj,\oo}/(\sum_{kl}
g_{kl,\oo}-i\oo C), \displaystyle\nonumber\\
g_{ij,\oo}&=&2\pi \nu_s \int d\e\Tr [\openone_i\openone_j -
\openone_j {\cal S}^\dagger(\e)\openone_i{\cal
S}(\e+\oo/2\pi)]\nonumber
\\
&&\times(f(\e)-f(\e+\oo/2\pi))/\oo,\,\,i,j,k,l=\overline{1,M},\label{eq:gs}
\end{eqnarray}
using summation over $k$ or $l$, $\sum_{k(l)=0}^M
g_{kl,\oo}^s=0$\,\cite{ButtikerPretre:1993PRL,BrouwerButtiker:1997}.
All probes except the current source 1 and sink 2, $I=I_1=-I_2$, are
set to voltages such that they draw no current. We shift all
voltages by $-V_{4\oo}$ and eliminate the 4th row and column from
$g^s$, invert the rest and obtain the resistance matrix $R$. It is
used to explore statistics of $\Rft=R_{31}-R_{32}$ and
$\Rtt=R_{11}-R_{12}-R_{21}+R_{22}$.

{\it DC setup, $\oo=0$.} First, let us consider dc transport and
ballistic QPCs at $T=0$, when any $\Tr\GG^n_j=N_j$ in
Eqs.\,(\ref{eq:acRtt},\ref{eq:Rmainac}) and the integral,
Eq.\,\,(\ref{eq:int}), is reduced to 1. In this limit the
measurements in a two-terminal dot are not affected by the Coulomb
interaction, and both $\la\Rtt\ra$ and its fluctuations are
diminished by the
TRS-breaking\,\cite{BrouwerLamacraftFlensberg:2005}. In a
multi-terminal setup, however, chaotic scattering results in a
random voltage drop between any pair of voltage probes. As a
consequence, the average $\Rft$ vanishes, $\la\Rft\ra=0$, and does
not have a WL contribution. Indeed, $\Rtt=(g_{33} g_{44}-g_{34}
g_{43})\cdot\mbox{det }R$ is symmetric with respect to
$1\leftrightarrow 2$, and its WL is the first correction due to
self-intersections of an electronic trajectory from 1 to 2. On the
other hand, $\Rft=(g_{31} g_{42}-g_{32} g_{41})\cdot\mbox{det
}R$\,\cite{Buttiker4terminal:1986} is anti-symmetric, and any loop
contributes equally to both terms. Consequently, WL corrections are
canceled, and they appear only if $\la\Rft\ra\neq 0$, \eg in a
quasi-1d disordered structure\,\cite{Ihn:2008} with a special probe
arrangement.

Similarly, the TRS-breaking does not affect the fluctuations of
$\Rft$ as shown by Eq. \,(\ref{eq:Rmainac}), in contrast with
Eq.\,(\ref{eq:acUCF}) for $\Rtt$. Indeed, inverting time we must
invert magnetic flux $\Phi$ and swap the current- with
voltage-probes. The ensuing symmetry $\Rtt(\Phi)=\Rtt(-\Phi)$ gives
$\DD\Rtt/\DD\Phi=0$ at $\Phi=0$, but there is no such a restriction
on $\Rft(\Phi)$. Fluctuations of $\Rtt(\Phi)$ are analogous to a
perturbation on a string with a free end, where the perturbation
maximum corresponds to the maximum in $\Rtt$-fluctuations at
$\Phi=0$. On the other hand, $\Rft(\Phi)$ is analogous to an {\it
infinite} string, so that zero field is as good as any other:
neither TRS-breaking effects like WL in Eq.\,(\ref{eq:acRtt}) or
$2/\b$-coefficient of Eq.\,(\ref{eq:acUCF}), nor "quenching"
(diminished sensitivity to $\Phi$ at $\Phi=0$) appear for $\Rft$ in
chaotic dots\,\cite{PolianskiIFE:2009}. In addition, in contrast to
the classical four-terminal scheme\,\cite{vanderpauw:1958}, the
properties of source and drain are never fully eliminated: a
measurement affects the sample and so the mesoscopic fluctuations in
$\Rft$ do depend on contacts. Importantly, in the experiment
\cite{Lerescu:2007} the voltage probes are invasive.

For the non-ideal contact coupling a tunneling limit for the voltage
probes is often taken, $\Tr\G_v\ll 1$; in the opposite limit,
$\Tr\G_v\gg 1$, UCF presents a complicated 6-th order polynomial of
$\GG_j$\,\cite{BrouwerBeenakker:1996}. In contrast, for a fixed set
$\{\GG_j\}$ the fluctuations of $\Rft$ given by
Eq.\,(\ref{eq:Rmainac}) can be expressed only in terms of
conductance $\Tr\GG_j$ and Fano factor $\F_j$, the ratio of shot
noise to current\,\cite{Blanter:2000}, of each QPC. The result of a
separate averaging over mesoscopic contact $j$,
$\Tr\GG_j^2/\Tr^2\GG_j\to(1-\la\F_j\ra)/\la\Tr\GG_j\ra$, is
proportional to the QPC resistance with a numerical prefactor: it
ranges from 1 for ballistic to 2/3 for a short diffusive wire, if
disorder is increased; for a wide tunneling QPC it is equal to 1 if
$\GG_j\ll 1$ are the same, and $1/2$ for a dirty
interface\,\cite{Schep:1997}. We observe that, generally, a poorer
coupling increases fluctuations of $\Rft$.

Often conductors have more contacts then needed for a four-terminal
measurement. A floating (unused) contact $f$ does not to draw any
current from the sample, but leads to
decoherence\,\cite{ButtikerPhysRevB:1986}, its coupling is related
to the inelastic scattering time
$\tau_{in}=h/\Delta\Tr\GG_f$\,\cite{Brouwer:1997}. Data are often
interpreted as an addition of several fictitious dephasing channels
to the real contacts. However, here it is important to distinguish
between a floating probe and a contact used for transport
measurements. Similarly to Ref.\,\onlinecite{Brouwer:1997}, we find
that a floating probe decreases fluctuations of $\Rft$ by increasing
the total coupling $\Tr\GG$ of the dot to reservoirs. Due to the
zero-dimensional geometry of the dot, they are independent of the
number of such probes, and only the total $\Tr\GG_f$ is relevant.
The voltage probes used to define $\Rft$ and floating probes are not
equivalent, an increase of the former decreases fluctuations
stronger. Indeed, in two measurements with unequal probes used as
voltage and floating ones and vice versa, the better the voltage
probe conducts the smaller the fluctuations of $\Rft$. Importantly,
if the dephasing probe model is used, dephasing channels should be
included not into the real, but into the floating probes.

{\it AC setup, $\oo\neq 0$.} Now we generalize our dc results to
finite frequency $\oo$ to consider dynamics of resistances. The
screened conductance $g^s_{\oo}$ in Eq.\,(\ref{eq:gs}) becomes
complex and depends not only on $\dwell$\,\cite{PieperPrice:1994},
but also on $\RC=C_\m/\Tr\GG$. Here $C_\m$ is the electro-chemical
capacitance, which relates the charge of the dot to the chemical
potentials, it is defined by the Coulomb interaction strength,
$1/C_\m\equiv 1/C+\Delta/\nu_s$\,\cite{ButtikerPretre:1993PRL}. A
frequency $\sim 1/\RC$, when the capacitor and the contacts conduct
similarly, is then a natural characteristic of a two-terminal
measurement. The finite-frequency correlations and WL of conductance
were considered only for ballistic QPCs at $T,\oo\to 0$. Quantum
effects in ac transport depended only on $\oo\dwell$ {\it at low
frequencies}\,\cite{BrouwerButtiker:1997}, but at higher $\oo$ they
start to depend on $\oo\RC$ as well. In contrast, for $\Rtt$ our
Eq.\,(\ref{eq:acRtt}) shows that {\it at any} frequency the WL is
independent of $\RC$; this is compatible with $\RC$-dependent WL in
{\it conductance} \,\cite{BrouwerButtiker:1997}, because $\RC$ is
eliminated by the matrix inversion. The reason is that WL only
increases the return probability, but does not redistribute charge.

Low-frequency conductance correlations can be generalized to
arbitrary frequencies $\oo,\oo'$ using parameter $z\equiv
-i\oo\RC/(1-i\oo\RC) $ and $z'$ introduced similarly, and
considering ballistic QPCs with $n_i=N_i/N$ for simplicity. For
$\b=2$ we find the correlations,
\begin{eqnarray}\label{eq:2Tcorr}
\frac{\la g_{ij,\oo}^sg_{kl,\oo'}^s\ra}{\nu_s^2 n_i n_j n_k
n_l}=\int d\tau\frac{(2\pi^2
T)^2e^{-\tau/\dwell}(e^{i\oo\tau}-1)(1-e^{i\oo'\tau})}
{\dwell\oo\oo'\sinh^2(2\pi^2 T\tau)}&&\nonumber \\
\times\left[
\frac{(\d_{ik}/n_k-1+zz')(\d_{jl}/n_l-1+zz')}{1-i\dwell(\oo+\oo')/2}
-\frac{\tau z^2 z'^2}{\dwell^3\oo\oo'}\right],&&
\end{eqnarray}
and for $\b=1$ the same r.h.s. with $k\leftrightarrow l$ should be
added. Results of Ref.\,\cite{BrouwerButtiker:1997} are reproduced
by differentiation $\DD_{\oo(\oo')}$ of Eq.\,(\ref{eq:2Tcorr})
 at $\oo,\oo'=0$. Using Eq.\,(\ref{eq:2Tcorr}) we find that fluctuations of
  $\Rtt$ in a dot with ballistic probes,
  $N=N_1+N_2+N_f$, given by Eq.\,(\ref{eq:acUCF}) manifestly depend on both $\dwell$ and $\RC$. Only
the four-probe scheme can show if $\oo\RC$ is an intrinsic scale of
the dot itself, or if it comes from the chosen two-terminal
measurement scheme. In other words, does the frequency-scale $1/\RC$
survive in resistance statistics if the role of contacts is
minimized? Equation (\ref{eq:Rmainac}) for $\Rft$ does not depend on
$\RC$, and we conclude that this scale is extrinsic and can be
understood as a contact property. Since Gaussian $\Rft$ is fully
characterized by Eqs.\,(\ref{eq:acRft},\ref{eq:Rmainac}), we
conclude that the quantum transport at {\it arbitrary} frequency
$\oo$ is insensitive to screening only in the four-terminal scheme.

{\it Experimental relevance.} Recently, Lerescu
\etal\,\cite{Lerescu:2007} measured statistics of $\Rft$ in
dc-biased ballistic dots and compared with numerical data for
$N_f=0$. The rms $\Rft$ qualitatively agreed with the numerics, but
was about 20 times smaller than they expected. These data at $N\gg
1$ indeed correspond to our predictions that $\la\Rft\ra=0$ and its
fluctuations are strongly diminished by an increased voltage probe
coupling. Their numerical data are well-fitted by
Eq.\,(\ref{eq:Rmainac}), which demonstrates a weak dependence on
$N_3-N_4$ and does not depend on $\b=1,2$. Reasons for the much
reduced experimental $\la\Rft^2\ra$ are strong decoherence, which
might depend on couplings, and possibly direct source-drain
trajectories, which result in voltage drop fluctuations smaller than
expected by RMT.

High-frequency experiments are more difficult. If a leakage current
through $C$ bypasses the drain, we have $I_{1\oo}\neq -I_{2\oo}$.
Resistance $\Rft$ relates the voltage drop $V_{34}$ to some linear
combination of these currents, and our $\Rft$ corresponds to
$(I_{1\oo}-I_{2\oo})/2$. If the voltage drop is measured via a
transmission line with length comparable to $2\pi c/\oo$, the
circuit should be also taken into account.  The phase shift of
transmitted voltage with respect to current source has a classical
component due to the circuit and a fluctuating part $\phi$ due the
dot, and we consider its mesoscopic distribution $P(\phi)$. If the
line impedance between the exits 3,4 (1,2) is large compared to
$\Rft(\Rtt)$, one has $\phi=\arg\Rft$. Using
Eqs.\,(\ref{eq:Rmainac},\ref{eq:int}) and $\la|\Rft|^2\ra$
calculated similarly, and normalizing $\oint d\phi P(\phi)=2\pi$, we
find
\begin{eqnarray}
P(\phi)&=&\frac{\sqrt{\int_{\oo 0}^2-|\int_{\oo\oo}|^2}}{\int_{\oo
0}-\mbox{Re }[\exp(-2i\phi)\int_{\oo\oo}]}=
\frac{\sqrt{a(a+1)}}{a+\sin^2(\phi-\phi_0)}.\label{eq:shape}
%
\end{eqnarray}
This $\pi$-periodic distribution is insensitive to the probe
properties and dephasing. The position $\phi_0$ and the height of
the distribution maximum are defined by Eq.\,(\ref{eq:shape}), and
Fig.\,\ref{FIG:setup}(b) presents $P(\phi)$ for low $T\ll\h/\dwell$.
The maximum $\sqrt 2/\oo\dwell$  at $\phi_0 =\oo\dwell$ for
$\oo\dwell\ll 1$ is sharp and the reactive part of $\Rft$ is small.
At
 $\oo\dwell\gg 1$ the peak at
$\phi_0=\pi/2-\log_2 \oo\dwell /(4\oo\dwell)$ is small due to weak
correlations between (re)active parts, and any phase $\phi$ is
possible. For high $T\gg\h\oo,\h/\dwell$ when RMT is still valid,
one has
$P(\phi)=\oo\dwell/[\sqrt{1+\oo^2\dwell^2}-\cos(2\phi-\arctan\oo\dwell)]$,
qualitatively similar to the low-$T$ result. We expect that the
universal distribution given by Eq.\,(\ref{eq:shape}) can be found
from phase measurements by varying either $\oo$ or $T$.

{\it Conclusions.} We discuss statistics of multi-terminal transport
through a well-conducting chaotic quantum dot with imperfect
coupling. Using RMT we find that four-probe resistance is unaffected
by weak localization and its fluctuations are governed only by the
dwell-time. Unlike two-probe transport, the four-terminal one is
insensitive to the Coulomb interactions at any frequency, suggesting
that the charge-relaxation time is connected to the measurement
scheme. The fluctuations are given by a single analytical expression
for arbitrary temperature, ac frequency, non-ideal coupling of the
contacts, and floating probes. We propose a universal mesoscopic
distribution for the phase of voltage transmitted through the dot.

{\it Acknowledgments. }We would like to thank K. Flensberg, J.
Gabelli, and C. M. Marcus for useful comments and discussion. MB is
supported by the Swiss NSF and MaNEP.



\end{document}